\documentclass[11pt]{article}

\usepackage{llncsdoc}

\usepackage{verbatim}
\usepackage{graphicx}
\usepackage{amsmath}
\usepackage{multicol}
\usepackage{booktabs}
\usepackage{subfig}
\usepackage{url}
\usepackage{algpseudocode}
\usepackage{algorithm}
\usepackage{xcolor}
\usepackage{pdflscape}
\usepackage[graphicx]{realboxes}
\usepackage{rotating}
\usepackage{pdfpages}

\usepackage[T1]{fontenc}
\usepackage[utf8]{inputenc}
\usepackage{authblk}

\algnewcommand{\LineComment}[1]{\State \(\triangleright\) #1}

\DeclareMathOperator*{\argmax}{arg\,max}

\begin{document}
\date{}



\title{
{\bf Learning Alternative Name Spellings}\\
\vspace{0.1in}
}

\author[1,2]{\rm Jeffrey Sukharev}
\author[1,3]{\rm Leonid Zhukov}
\author[1]{\rm Alexandrin Popescul}
\affil[1]{Ancestry.com, San Francisco, USA}
\affil[2]{University of California, Davis, USA}
\affil[3]{Higher School of Economics, Moscow, Russia}
\affil[ ]{\textit {\{jsukharev,lzhukov,apopescul\}@ancestry.com}}

%

\maketitle

\begin{abstract}

Name matching is a key component of systems for entity resolution or record linkage. Alternative spellings of the same names are a common occurrence in many applications. We use the largest collection of genealogy person records in the world together with user search query logs to build name matching models. The procedure for building a crowd-sourced training set is outlined together with the presentation of our method.
We cast the problem of learning alternative spellings as a machine translation problem at the character level. We use information retrieval evaluation methodology to show that this method substantially outperforms on our data a number of standard well known phonetic and string similarity methods in terms of precision and recall. Additionally, we rigorously compare the performance of standard methods when compared with each other. Our result can lead to a significant practical impact in entity resolution applications.
\end{abstract}

\section{Introduction}

A person's name, especially the family name, is the key field used in identifying person's records in databases. Software tools and applications designed for entity resolution of a person's records usually rely on the person's name as a primary identification field. In particular, genealogy services provide user access to a person's record databases and facilitate search for a user's ancestors/relatives and other person/records of interest. A persons's records indicate some type of a life event including birth, death, marriage or relocation. Typically, records are indexed by some unique identifier and can also be searched by a combination of last/first names, geographical locations and event dates.
 Searching a record database is complicated by the user not knowing the exact spelling of the name in the record they are searching for. This task becomes even harder since databases often contain alternate spellings referring to the same person.  This transpires due to many factors including optical character recognition errors when scanning the records, errors/misspelling of names in the records themselves, and name transliterations. For instance a common last name ``Shepard'' has been also commonly spelled as ``Shepherd'', ``Sheppard'', ``Shephard'', ``Shepperd'', ``Sheperd''. Clearly, having methods that would provide users and the search engines with a list of highly credible misspelling of the query name would significantly improve the search results.

Knowing how to misspell names to find records of interest has always been a part of a professional genealogist's domain expertise. In this paper we try to bridge this gap and bring this power to the average user by employing data-driven methods that rely on our unique data set\footnote{Data is available for research purposes from the authors}. Through Ancestry.com we have access to the world's largest genealogy data repository. The main function of a genealogy service is to facilitate discovery of relevant person's records and the construction of family trees. Tree construction involves user attaching relevant scanned/digitized records, found in Ancestry.com databases, to user-generated tree node. Having records attached to individual tree nodes affords us an opportunity to collect misspellings of names. By leveraging user-provided links between individual user family tree nodes and scanned attached records, we generated a "labeled" dataset of name pairs where the left side of each pair comes from user supplied person names and the right part of the pair comes from the attached record name field. All user-identifying information except for last name pairs is discarded from the final dataset. We filter and pre-process this list of name pairs and use it to train a model using standard machine translation methods. We will go into more details on data pre-processing in Section 5. Additionally, we generated another dataset from company search logs. Often, users modify a previous search query in hope of getting better results. These user-driven modifications are called query reformulations. By identifying logged-in user sessions and extracting names from user queries in sequential order from the same session in a specified time interval and using our assumption that users frequently search for variations of the same name we have been able to accumulate a large number of name pairs that could also be used as a training/testing data for our models.

As a result of our experiments we produce ranked candidate variant spellings for each query name. In addition to providing the translation model we also propose a methodology, adapted from the information retrieval community, for evaluating of the final candidate list and for comparing it with other methods. 

In the results section we will show that our methods perform significantly better than other state-of-the-art methods in terms of precision and recall in identifying a  quality lists of alternative name spellings.

The remainder of the paper is organized as following. In section 2 we discuss the numerous previous works in related fields. A detailed description of our training data is given in Section 3. In Section 4 we outline the machine translation method used in training our model. We then discuss our results and present comparisons with other methods in Section 5 and conclude in Section 6.
\section{Previous work }

The classic reference in the field of record linkage is  a paper by Fellegi and Sunter \cite{fellegi1969} published in 1969.
In their work the authors have carefully presented the theory of record matching. They defined the terms of positive disposition (link and non-link) and negative disposition (possible link) and showed that the optimal record matching linkage rule would minimize the possibility of failing to make a positive disposition for fixed levels of errors. 
Since this seminal work there has been proliferation of work in this area. In the interest of brevity we direct the reader to the outstanding 2006 survey paper by Winkler \cite{Winkler06} and to the comprehensive work by Christen \cite{data:matching} published just recently.

With the explosive growth of data coming from web applications it is becoming imperative to discover the best methods for record matching in terms of accuracy and speed. Historically, methods focusing on name matching could be separated into two classes: sequential character methods and bag-of-words methods \cite{MoreauYC08}. 

\subsection{Sequential Character methods}
Phonetic similarity methods are an important category of sequential character methods. 
The key concept of phonetic methods is to map n-grams of characters into phonetic equivalents. The output of using these methods on string pairs is a binary decision and not a degree of similarity. The best-known method from this class is Soundex \cite{Russell18}. Over the years a numerous improvements of this approach have been made. In particular some of them had to do with accommodating non-English names. Popular methods include Soundex \cite{Russell18}, Double Metaphon \cite{philips90}, and NYSIIS \cite{Taft70}. While these methods proved to be useful in improving performance in data matching applications they do not solve the problem of relevance ranking of alternative spellings, which is of great importance for search engines when considering using alternative name spellings for query expansion.

Another important category of sequential character methods often used in conjunction with phonetic methods is the class of static string similarity measures.
Similarity method based on edit distance (the Levenshtein distance, as it is also known \cite{Levenshtein66}) is the most well-known method of this type. The edit distance between strings $s$ and $t$ is the cost of the optimal shortest sequence of edit operations (substitute, add, delete) that converts $s$ to $t$. For instance, the mapping of $s$ = ''Johnson'' to $t$ = ''Johnston'' results in one addition of letter "t" and hence, results in a distance of one. 
Other common similarity measures include the Jaro \cite{jaro89} method which takes into account the number
and order of common characters between two strings and the Jaro-Winkler \cite{winkler90}  method which extends Jaro by accounting for the common prefixes in both strings \cite{bilenko03}, \cite{data:matching}. 
The static similarity measures described above, while useful in measuring similarity and ranking alternative spellings, are not capable of generating alternative spellings. This capability is typically absent from all methods that do not take a  dataset's statistical information into account.

In 2013 Bradford \cite{Bradford13} published a paper dealing with alternative name spelling generation. He used latent semantic indexing that uses Singular Value Decomposition (SVD) method to identify patterns in the relationships between the terms in unstructured collection of texts.

Because of the difficulty associated with obtaining the experimental data many researchers 
 build their own synthetic datasets by utilizing specialized tools for mining the web and extracting words that appear to be last names. The resulting names are used in forming artificial pairs using one or more similarity measures (typically based on edit distance). Another popular alternative is to hire human annotators who create last name pairs based on their knowledge of name misspelling. Both of these methods may introduce bias. 

Our data is being produced by millions of users who act as human annotators and who should be experts in their own genealogy and are motivated to build quality content. Due to the nature of our dataset we can extract best pairs using frequency statistics. We will go into more detail about our filtering process later in this paper.
Having frequency information allows us to assemble realistic distribution of name pairs and helps in training more accurate models of alternative name spellings.

\subsection{Bag-of-words methods}
Bag-of-words methods typically represent strings as a set of words or word n-grams. There were numerous studies published on the topic of applying bags of words to record linkage over the last decade \cite{MoreauYC08}. Cosine similarity of term frequency inverse document frequency (TFIDF)  weighted vectors is one of the most popular methods of this type. Typical vectors consist of individual words or n-grams. The main shortcoming of cosine similarity TFIDF is that this method requires exact matches between fields. To alleviate this issue cosine similarity SoftTFIDF was introduced by Cohen et. al. \cite{Cohen03acomparison}. In addition to counting identical fields occurring in both vectors SoftTFIDF compares and keeps track of "similar" fields in both vectors. Bilenko et. al. \cite{bilenko03} showed how machine learning methods could be successfully employed for learning the combined field similarity. They trained an SVM classifier using feature vectors, and then applied the learned classifier's confidence in the match as a class score. 
In this paper we do not consider these approaches because we primarily work with single word last names and bag-of-words methods are more suited for finding similarity between multi-field records.

\subsection{Spelling correction and Machine Translation literature}
In the 1990 Kernighan et. al. \cite{kernighan1990spelling} in their short paper proposed a method for spelling corrections based on noisy channels. The same formulation would latter be used in machine translation field. The basic idea was to find best possible correction by optimizing the product of language model ( a prior probability of letters/phrases/words  in a given language) and correction model (likelihood of one word being spelled as another).  
For comprehensive survey of spelling correction methods the reader should look at the
excellent chapter on this topic at Jurafsky and Martin 2008 Speech and Language Processing book. \cite{jurafsky2008speech}

In the last several decades machine translation methods have gained significant traction and recently found their way into the problem of name matching. In 2007 Bhagat et. al. \cite{BhagatH07} implemented a transducer based method for finding alternative name spellings by employing a graphemes-to-phonemes framework. Their method involved running EM (expectation maximization) algorithm, first presented by Dempster \cite{dempster1977}, to align text from the CMU dictionary with their phoneme sequence equivalents. 
Next, they built a character language model of phoneme trigrams using the same CMU dictionary phonemes. Their training set was mined from the web.
 Using both-ways translation models and language models, the authors were able to generate alternative phoneme sequences (pronunciations), given a character string name, and then each of these sequences was converted into an alternative character sequence \cite{Pfeifer96}.
	
	In 1996 Ristad and Yianilos \cite{Ristad96} presented an interesting solution where they learned the cost of edit distance operations, which are normally all set to one in static edit distance algorithms. The authors used expectation maximization algorithm for training. Their model resulted in the form a transducer. 
	Bilenko et. al. \cite{BilenkoKDD03} improved on Ristad and Yianilos's learned edit distance model by including affine gaps. They also presented a learned string
similarity measure based on unordered bags of words, using SVM for training.
	McCallum et. al. \cite{McCallumBP05} in 2005 approached the same problem from the different angle. Instead of using generative models like \cite{Ristad96} and \cite{BilenkoKDD03} they have used discriminative method, conditional random fields (CRF), where they have been able to use both positive and negative string pairs for training.

\section{Datasets}

Ancestry.com has over the years accumulated over 11 billion records and over 40 million personal family trees \cite{acom12}. Most of the records in  the Ancestry.com database originate from the Western European countries, United States, Canada, Australia, and New Zealand.
 Scanned collections of census data, and personal public and private documents uploaded by company users comprise the bulk of Ancestry.com datasets. 
One of the key features of the Ancestry.com web site is the facility for building personal family trees with an option for searching and attaching relevant documents from record databases to the relevant parts of family trees. For example if a family tree contains a node for a person with the name John Smith it would be often accompanied by the birth record, relocation record, marriage record and other records discovered by the owner of the family tree. Since most of the nodes in the deep family trees involve persons who are no longer living, death records can often be discovered and attached to the appropriate nodes.

 This linkage between user-specified family tree nodes and the official records present us with a unique opportunity to assemble a parallel corpus of pairs of names, hand-labeled by the users themselves. In the past researchers working on name matching problem were forced to assemble their training datasets by employing text mining techniques. Very often a specific method was needed for identifying names in a given text and then edit distance measure was used to find a list of misspelling candidates. Additionally, in some studies, a small number of dedicated human labelers provided additional level of confidence. These methods would inevitably lead to bias. We believe that our user-labeled dataset contains significantly less bias than previously used training datasets. 

Due to the availability of the ``labeled'' dataset in the Ancestry.com we have a more direct way of generating training data. From the begininning we realized that we could not employ standard supervised machine learning methods for finding alternative name spellings since that would require us to collect positive and negative training sets. While it would have been possible to mine positive sets from user-labeled data, defining the process generating realistic negative examples is ambiguous at best. This would require us finding name pairs that would not be alternative spellings of each other with a high degree of confidence. Even through it may seem doable at first glance this a very tricky proposition. First of all how would we choose each pair item? What is the distribution of negative pairs? We only have user labels for positive pairs, but not having user label for a name pair does not necessarily mean that the pair is negative. Not having any other alternatives we would have to bias our negative set to some kind of similarity measure like the Levenshtein method and this would force us to arbitrary select a threshold that would distinguish negative pairs from positive pairs. However, besides introducing bias this method would make us miss numerous negative pairs  which would have high similarity values but would not constitute a positive common misspellings.
Due to having this obstacle in front of us, we turned toward machine translation methods because only a parallel corpus was needed to train the translation model.

Given the way Ancestry.com users interact with the genealogy service, we isolated two separate ways of collecting parallel corpus data that would later be used for training translation models and for testing. We felt that having two completely different underlying processes for generating our datasets would strengthen our case if we arrived at similar conclussions.

The first process of assembling a parallel corpus consists of collecting all directed pairs of names drawn from anonymized (striped from all user identifying information except last names) user tree nodes and their attached anonymized records. We chose pair direction as following: last names on the left come from tree nodes and last names on the right come from the records. Since last names in records and tree nodes have different distribution taking directionality into account is important when choosing the training set of pairs.  A number of filtering steps have been applied in order to de-noise the datasets and will be discussed in more detail in the later sections. The pairs are directed which implies that a pair ``Johansson'' - ``Johanson'' would be different from the reverse pair ``Johanson''-``Johansson''. This would manifest in separate co-occurrence count for each pair. 

The second process for building a parallel corpus involves using recorded user search queries. Since the Ancestry.com search query form asks the user for specific fields when searching for trees or records, we have been able to extract user queries containing names from a search log.
By grouping users by their loginname, sorting the queries in chronological order, and fixing the time interval at a generous $30$ minutes, we have been able to extract directed pairs of names that users use in their search queries. Our build-in assumption is that frequently users do not find what they are looking for on their first attempt and if that is the case they try again. The resulting data set is also noisy and requires extensive filtering before being used as a training set. Each pair has a direction from an older name spelling to a newer reformulation. For example if a user A searches for name ``Shephard'' at time $t_0$ and then searches for name ``Shepperd'' at time $t_1$ where $t_1 - t_0 < 30$ then the resulting pair will be: ``Shephard'' - ``Shepperd'' and not the other way around.

Table \ref{table:names} provides an illustration of a sample of ``records'' dataset grouped by Levenshtein edit distance and sorted by co-occurrence count. The distribution of values of edit distances between names in each pair and types of individual edit operations needed to transform left-hand member of a pair into a right-hand member are shown on Tables \ref{table:ed} and \ref{table:ed_pct} for each dataset. 
We also demonstrate the breakdown of unique last names by their country of origin on Tables 4 and 5 for both datasets. Country of origin information was gathered from person tree nodes. Each person's node contains person's place of birth in addition to first and last names. The most common country of birth was selected as a name's country of origin. Only the ``Old World'' countries were chosen in order to avoid mixing names from different regions which are present in the ``New World'' countries.

\begin{center}
\begin{landscape}
\begin{table*}[ht]
\centering
\begin{tabular}{ |l|l|l|l|l|l|l|l|l|l| }
\hline\hline
Levenshtein edit distance & name\#1 & name\#2 & cooccurrence & count\#1 & count\#2 & Jaro-Winkler & Jaro & Jaccard  \\ [0.5ex] 
\hline
 
1 & clark & clarke & 139024 & 1168804 & 335902  & 0.922  & 0.889  & 0.102 \\
 & bailey & baily & 89910 & 725361 & 123012  & 0.922  & 0.889  & 0.119 \\
 & parrish & parish & 77529 & 179308 & 138774  & 0.933  & 0.905  & 0.322 \\
\hline
2 & seymour & seymore & 15583 & 90071 & 24127  & 0.907  & 0.810  & 0.158 \\
 & schumacher & schumaker & 6013 & 52769 & 12867  & 0.884  & 0.793  & 0.101 \\
 & bohannon & bohanan & 5902 & 44770 & 16252  & 0.854  & 0.738  & 0.107 \\
\hline
3 & arsenault & arseneau & 1489 & 11455 & 4305  & 0.838  & 0.769  & 0.104 \\
 & blackshear & blackshire & 1269 & 9556 & 3049  & 0.884  & 0.793  & 0.112 \\
 & grimwade & greenwade & 781 & 1886 & 2480  & 0.764  & 0.611  & 0.218 \\
\hline
4 & sumarlidasson & somerledsson & 671 & 674 & 1526  & 0.752  & 0.628  & 0.439 \\
 & riedmueller & reidmiller & 143 & 438 & 556  & 0.736  & 0.664  & 0.168 \\
 & braunberger & bramberg & 131 & 624 & 277  & 0.802  & 0.674  & 0.170 \\
\hline
\end{tabular}
\caption{Name pairs and statistics}
\label{table:names}
\end{table*}

\end{landscape}
\end{center}
\section{Methods}

The problem of finding best alternative name spellings given a source name can be posed as maximization of conditional probability $P(t_{name}|s_{name})$ where $t_{name}$ is a target name and $s_{name}$ is a source name. Following the traditions of statistical machine translation methods \cite{Brown1990} this probability can be expressed using Bayes' rule as 
\[
P(t_{name}|s_{name})  =  \frac{P(s_{name}|t_{name}) *  P(t_{name})}{P(s_{name})}
\]
where $P(t_{name})$ is a ''name model'' (corresponds to language model in machine translation literature) and describes frequencies of particular name/language constructs. 
$P(s_{name})$ is a probability of a source name. $P(s_{name} | t_{name})$ corresponds to alignment model.

''Name model'' can be estimated using character n-grams language model representation by finding the probabilities using the chain rule \cite{shannon48}):
\[
P (c_1 c_2 ... c_m) = \prod_{i=1}^{m} P (c_i|c_{max(1,i-(n-1))},...,c_{max(1,i-1)})
\]
where $c_i$ is an $i^{th}$ character in the sequence of characters that comprise a name of length $m$. $n$-gram model computes a probability of a character sequence where each subsequent character depends on $n-1$ previous characters in the sequence. 

\begin{figure}
\centering
\includegraphics[width=7.2cm]{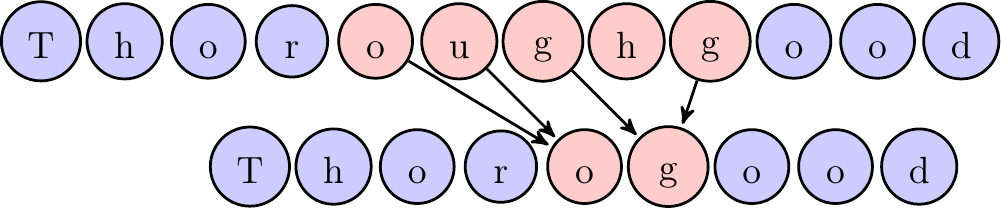}
\caption{Machine translation: alignment; the source name ``Thoroughgood'' and the target name ``Thorogood''. Arrows and red circles represent phrase alignment rules learned as a result of the training stage. }
\label{fig:align}
\end{figure}

An ``alignment model'' is used in generating translational correspondences between names in our context and it can be best described by an example shown on Figure \ref{fig:align}. Here the name ``Thorogood'' is aligned with the name ``Thoroughgood''.  Looking at the Figure \ref{fig:align} we can clearly see that second occurrence of letter 'o' in ``Thorogood'' alignes with two letters ('o' and 'u') in ``Thoroughgood'', similar situation happens with the last letter 'g' in ``Thorogood'' which gets aligned with 2 letters 'g' in ``Thoroughgood''. Other letters in ``Thorogood'' aligned 1-to-1 with letters in ``Thoroughgood''. Letter 'h' in ``Thoroughgood'' does not align with anything in ``Thorogood''. 
Estimating ``alignment model'' results in generation alignement rules such as the ones that we just presented.

We are not only interested in the best alternative spelling given by\\ $\arg \max\limits_{t_{name}} P(t_{name}|s_{name})$, but also in the ranked list of best suggestions, that can be computed from the same distribution by sorting probabilities in decreasing order:

\[
\argmax^{K}_{t_{name}} P(t_{name} | s_{name})  =  \argmax^{K}_{t_{name}} P(t_{name}) * P(s_{name} | t_{name}) 
\]



where $\arg \max\limits^{K}_{t_{name}}$ represents operator that finds top $K$ $t_{name}$s that maximize $P(t_{name}|s_{name})$. Finding $P(t_{name} | s_{name})$ accurately without using the equation above would be challenging. However, using Baye's rule and breaking down $P(t_{name} | s_{name})$ into language model $P(t_{name})$ and alignment model $P(s_{name} | t_{name})$ allows us to get a theoretically good translation even if underlying probabilities are not that accurate \cite{Brown1990}. $P(s_{name})$ is fixed and does not depend on the optimization variable $t_name$ and hence, will not influence the outcome and can be discarded.

To find probability values corresponding ``name model'' and ``alignment model'' we will be using tools developed by machine translation community, replacing sentences with names and words with characters.

For training of our language and alignment models we have chosen the Moses software package which is a widely known open-source statistical machine translation software package \cite{Koehn2007}.
Moses is a package that contains various tools needed in translation process.
Typically, translation software deals with words in a sentence as primary tokens, since we compare individual last names we had to transform our input to a format recognizable by Moses while also maintaining characters as primary tokens. In our case single words become sentences and characters become words in the sentence.   

When using the Moses software package we chose to use Moses' Baseline System training pipeline.
It includes several stages:

\begin{enumerate}
\item Preparing the dataset: tokenization, truecasing and cleaning. Tokenization involves including spaces between every character. Truecasing and cleaning deals with lowercasing each string and removing all non-alphabetic characters among other things.
\item Language model training. A language model is a set of statistics generated for an  n-gram representation built with the target language. We used IRSTLM \cite{Federico08}, a statistical language model tool for this purpose.
As a result we generated 2-gram through 6-gram language models (6 was the maximum possible). This step adds ''sentence'' boundary (''word'' boundary here) symbols and, also as in the Baseline System, uses improved Kneser-Ney smoothing. We follow a common practice in machine translation where all examples of the target language, and not only forms present in parallel corpus translation pairs, are used to construct a language model. It is, therefore, based on a larger data set, and can lead to an improved translation quality. In our experiments with search logs and tree attachment datasets we used their respective lists of 250,000 most frequent surname forms for language model estimation.
\item
Alignment model building: Moses uses the GIZA++ package for statistical character-alignment \cite{och03:asc}  character (Word)-alignment tools typically implement one of Brown's IBM generative models \cite{Brown93} that are being used for determining translation rules for source language to the target language (including fertility rules: maximum number of target characters generated from one source character and so on) We created alignment model, for each of the 2-gram through 6-gram language models created in the previous step. As in the Baseline System, the ''-alignment'' option was set to ''grow-diag-final-and'' and
the ''-reordering'' option was set to ''msd-bidirectional-fe''          

\item
Testing. We tested decoding on test folds in a batch mode with an option ''-n-best-list'' to give top 1000 distinct translations. This value was chosen large to well represent the high recall area on respective precision-recall curves.
It is possible that using different Moses configuration could give even more accurate results.

\end{enumerate}

We basically followed the Baseline System with the exception of tuning the phase and replacing our source and target languages with sequences of characters and instead of sequences of words. The tuning phase consists of steps optimizing the default model weights used in the training phase. We have omitted this phase because based on our initial tests, it didn't give immediate accuracy improvements on our datasets and it is relatively slow. 
\begin{center}
\begin{table}

\centering
\begin{tabular}{ |l|l|l| }
\hline
Edit Distance & ``Search'' \# of pairs & ``Records'' \# of pairs \\ [0.5ex] 
\hline
 
1 & 10894 & 21819 \\
2 & 1312  & 2560 \\
3 & 155   & 258 \\
4 & 32    & 70 \\
5 & 15    & 58 \\
6 & 20    & 66 \\
7 & 30    & 68 \\
8-11 & 42    & 99 \\
\hline
\end{tabular}

\caption{Edit distance distribution}
\label{table:ed}
\end{table}
\end{center}

\begin{table}

\centering
\begin{tabular}{ |l|l|l| }
\hline
Operations & ``Search'' ops type \% & ``Records'' ops type \% \\ [0.8ex] 
\hline
 
deletes &  32.18 \% & 38.18 \% \\
inserts & 33.91 \%  & 20.65 \% \\
replaces  & 33.91 \%  & 41.47 \% \\
\hline
\end{tabular}

\caption{Distribution of operations among name pairs separated by edit distance 1}
\label{table:ed_pct}
\end{table}

\section{Results}
\subsection{Data Preparation}
Since we dealt with user-generated data we had to devise an algorithm for treating the data and generating a high confidence training set. We outlined the following procedure:
\begin{enumerate}
\item
Initially, a ''universe'' of names was defined. All names in tests and training sequences came from this "universe". A set of names was selected by taking top 250,000 most frequent names from both datasets (``search'' and ``records'').
\item
For each pair selected using procedure outlined in Section 3 we made sure that each name comes from our set of high-frequency names.
This step resulted in $12,855,829$ pairs in the ``search'' dataset and $51,744,673$ pairs in the ``records'' dataset.
\item In order to de-noise the name pairs we selected the top 500k/250k pairs by co-occurrence for the ``records'' and ``search'' datasets respectively.

\item The remaining pairs were passed through the Jaccard index filter $J$:
\[
J(A,B) =  |(A \cap B)| / |(A \cup B)|
\]
where $A$ is a set of users linked with left name from a name pair and $B$ is set of users linked with the right names from a name pair. Users are identified by either their login session (``search'' dataset) or by userid (``records'' dataset). The reason that users where used in calculating Jaccard instead of just using co-occurrence counts and marginal counts has to do with preventing a few highly active users from skewing the results. This filter was used to remove name pairs that would be likely to co-occur by chance due to high frequency of each individual name involved in a pair. For instance, ``Smith''-``Williams'' pair would be filtered out.
After filtering we were left with 25k "record" and 12.5k ``search'' name pairs.

\item
In the final step we estimated the rate of "obvious" false positives based on manual checks and similarity measures cross checking.
Looking at random samples stratified by edit distance we manually evaluated these samples to estimate the false positive percentage.
We estimated that the rate of obvious false positives is 1.5\% in ``search'' dataset and 1.4\% in ``records'' dataset. We specifically avoided using  string based similarity criteria when defining parallel corpus to prevent introducing bias. In principle, extra filters can be applied to training sets. 
\end{enumerate}

\subsection{Experiments and results}
Comparing phonetic methods with similarity measures and with machine translation methods is not straightforward. Phonetic methods only allow for binary responses (match or mismatch) when applied to name pairs. Therefore, it is impossible to rank positive matches without introducing additional ranking criteria.
Our machine translation (MT) method produces a score that we use in ranking. Similarity methods produce a similarity value that is also used in ranking. To get a meaningful comparison of these methods with phonetic methods we had to make use of statistics that we gathered while processing datasets. Additionally, we devised a unified methodology that could be applied to all listed method types.

In all our experiments we used 10-fold cross validation for evaluating how the results of predictions generalize to previously unseen data.

We randomly divided each dataset (``search'' and ``records'') into ten folds. We train on 9 folds, then test on the remaining 1. This process was repeated 10 times for each test fold. Training folds were used to train the MT models. The same test folds were used to test all methods, including MT generated models, phonetic methods and similarity measures. 

Generating results involves building a consistent metric that can be plotted and compared between different methods.
We adapted a standard information retrieval performance metric: precision and recall. 
\[
Precision = \frac{TP}{TP+FP}
\]

\[
Recall = \frac{TP}{TP+FN}
\]
where $TP$ are true positives, $FN$ are false negatives and $FP$ are false negatives.
The methods with larger precision and recall are superior. 

Each test fold contains a source name and one or more target names associated with each source name. Each of our methods for each source name would produce its own list of target names. Since the number of suggested target names (or alternative spellings) can be large we needed to find a suitable method for ranking target names. For all target names  for position/rank $i$  in the range from $1$ to $N$ corresponding $recall_i$ and $precision_i$ are calculated. So, we had to agree on what precisely we mean by rank for phonetic methods, similarity methods and machine translation methods. 

We decided to view ranking as the product of 
\[
rank(s,t) = alignmentScore(s,t) * languageModelScore(t)
\]
For the machine translation method (generated using the Moses software library) we used model-applier output scores which already contain the product of language model score and alignment score.
For machine translation where character is a \emph{word}:  
\[
rank(s,t) = mosesScore(s, t)
\]
For phonetic algorithms $languageModelScore(t)$ is the frequency of a name in the dataset ($freq$). 
\[  
rank(s,t) = hasSameCode(s, t) * freq(t)
\]
where $hasSameCode(s, t) \to \{0,1\}$ and $freq(t)$ represents the frequency of name $t$ in the dataset.

For similarity measures we also used name frequency, but we had to experimentally find a suitable exponential constant $\gamma$ to avoid over-penalizing low-frequency names. 
\[ 
rank(s,t) = sim(s, t) * freq(t)^\gamma
\]
where $sim(s,t)\to [0,1]$ represents the floating point similarity values and $\gamma$ is the exponential constant used to control the frequency values. We used $\gamma$ value $0.001$.

After saving precomputed sorted (according to ranking) lists of alternative name spellings for each method (phonetic, similarity, MT methods) we computed Precision and Recall values for each position from $1$ to $N$ separately for each test fold. 

After producing 10 precision-recall curves for each method we needed to find a suitable way to visualize confidence in our results without actually drawing 10 curves per method.

Inspired by the work of \cite{Macskassy04} we designed our own methodology for robust statistical comparisons of our precision-recall curves. Using our ten folds we evaluated confidence bands for each method. Assuming test examples are drawn from the same, fixed, multivariate normal distribution, the expectation is that the model's precision-recall curves will fall within the bands with probability of at least $1-\delta$ where $\delta$ represents the significance level. We need to find the standard deviation of the sample which is the degree to which individual points within the sample differ from the sample mean. 

The density contours of  multivariate normal distribution of precision and recall pairs are ellipses centered at the mean of the sample. The eigenvectors of the covariance matrix $\Sigma$ are used as directions of the principal axes of the Gaussian ellipses \cite{Hansen05thecma}.
For our collection of $2D$ precision/recall pairs $X = (X_1,X_2)$
\[
\Sigma_{ij} = E[(X_i-\mu_i)(X_j-\mu_j)]
\]

The average values of ten points $\mu_1$ and $\mu_2$ have given us centroid curve for each method and the center of density contours.

\begin{figure*}
\centering
\subfigure[Zoom-out: The borders of Gaussian regions  $2.447*\sigma$ used to generate confidence regions]{\includegraphics[width=8.5cm]{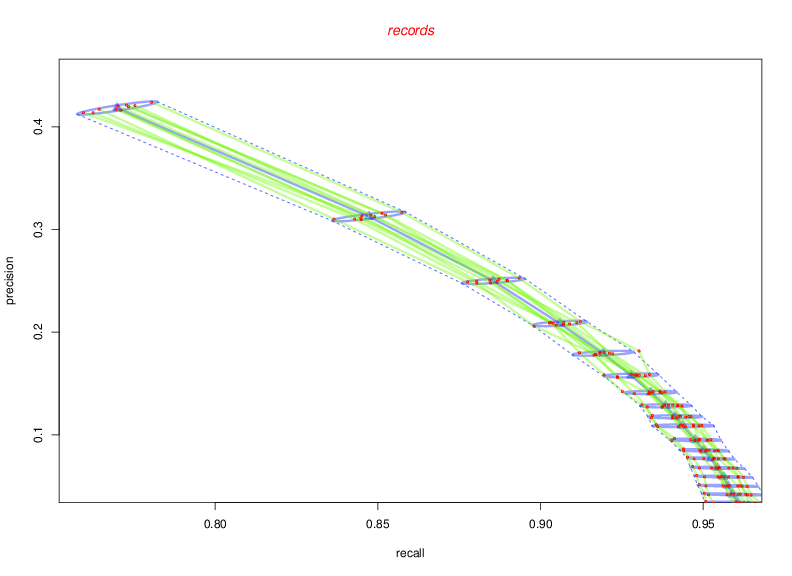}}

\subfigure[Zoom-in: The borders of Gaussian regions  $2.447*\sigma$ used to generate confidence regions]{\includegraphics[width=8.5cm]{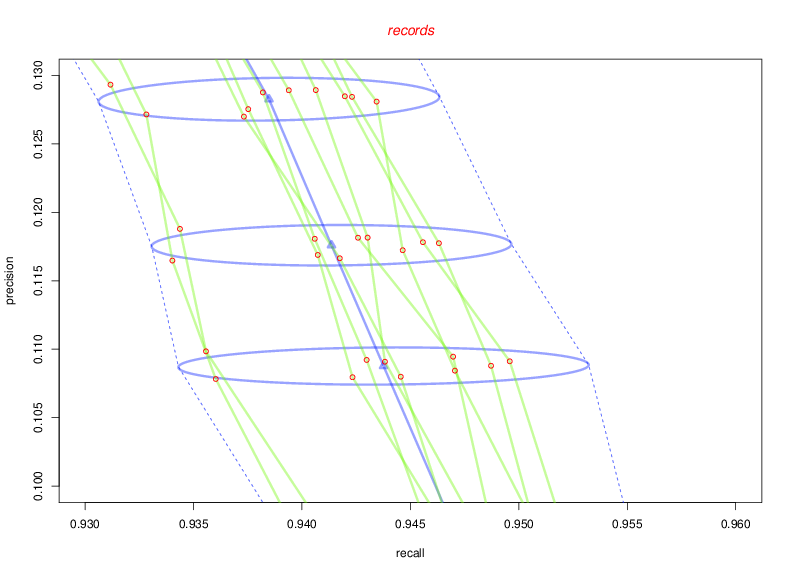}}
\hfill
\caption{Confidence bands for Moses produced 5-gram MT method. Green lines show $10$ folds and small red circles indicate ranking position for a given curve on the precision/recall chart.
For each position there are as many points as there are folds and the coordinates for these positions, according to our assumption, are normally distributed for both precision and recall. After computing elliptical density contours, shown in light blue, we connected the endpoints of the longest principle axis of the neighboring ellipses to produce confidence bands (indicated by dashed light blue lines). }
\label{fig:bands}
\end{figure*}

The standard deviation for each vector direction is found by taking the Cholesky decomposition of the covariance matrix and using the resulting matrix for generating elliptical contours of a two dimensional normal distribution. To capture the $95\%$ confidence level in $2D$ we need to multiply each $\sigma$ by the multiplier. 
We get the squared $\sigma$-multiplier value from the Chi Square Distribution ($\chi^{2}$)  table for $2$ degrees of freedom where the Chi Square Distribution is the distribution of the sum of squared independent standard normal variables. $\sigma$ multiplier equals to $2.447$ in this case.
See Figure \ref{fig:bands} for a visualization and further explanation of confidence bands.

The resulting bands are formed by connecting by line segment endpoints of the longest principle axis of each ellipse with its corresponding neighbor ellipses.  The resulting bands give us a visual cue regarding the variance of precision/recall (PR) curves produced for different data test folds. Also, the resulting bands have at least $95\%$ confidence level because data points that may not be captured be ellipses may still end up inside the bands between ellipses and since the ellipses are already at $95\%$ confidence level this implies that the bands will have a higher confidence level.

We ran 70 experiments on phonetic methods. Seven commonly used phonetic methods were selected for testing and these methods were applied on the same ten test folds.
90 experiments were conducted with distance metrics methods (Jaro, Levenshtein, Winkler-Jaro). We experimented with three values when choosing suitable $\gamma$ parameter for distance measurement methods ranking.
Our results indicate general consistency when using test data from both datasets (``search'' and ``records''). NYSIIS phonetic method, first introduced by Taft in 1970 \cite{Taft70} significantly outperforms other phonetic methods. Phonex method appears to be the weakest performer of the phonetic methods we have looked at.
Other phonetic methods lie in the middle and their confidence bands overlap. Because of the overlapping regions we cannot definitively rank the performance of these methods.

\begin{figure*}
\centering
\subfigure[``Search'' dataset]{\includegraphics[width=8.5cm]{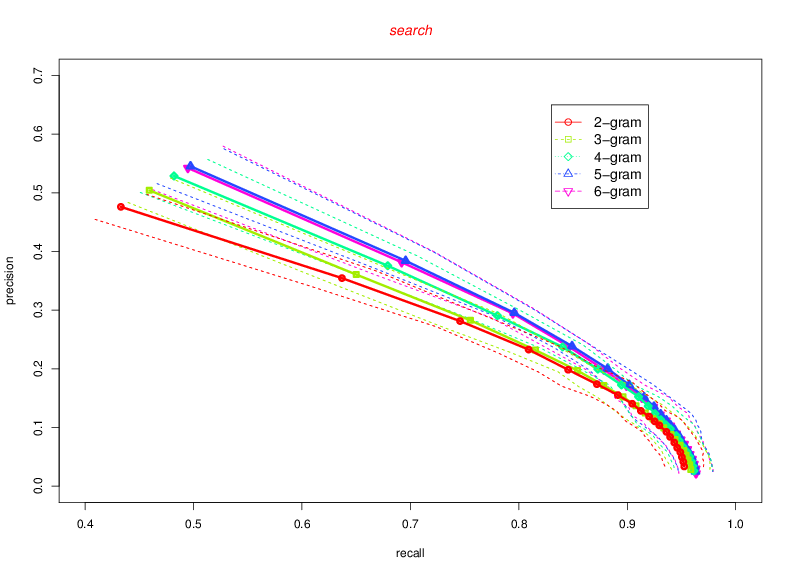}}

\subfigure[``Records'' dataset]{\includegraphics[width=8.5cm]{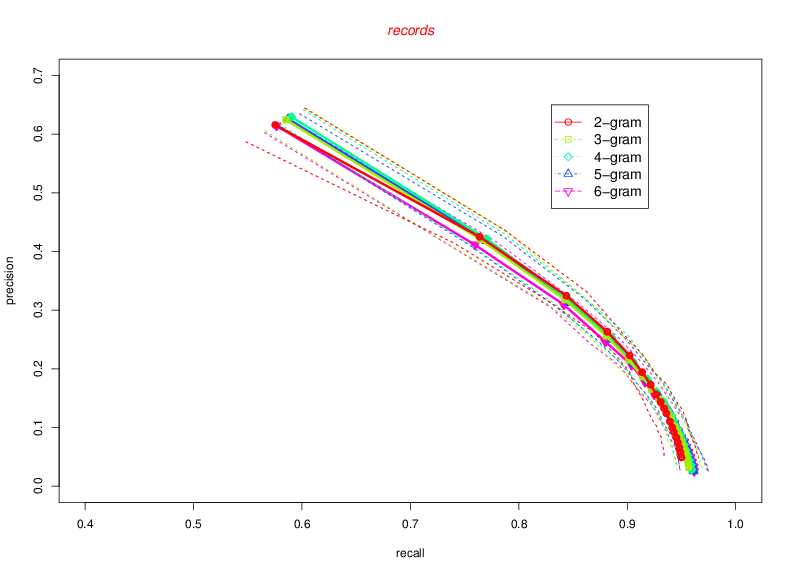}}

\caption{MT model experiments. In ``search'' dataset 5-gram/6-gram methods appear to perform slightly better than other MT models. On the other hand, in ``Records'' dataset almost all methods confidence bands overlap and it is hard to make a definitive statement as to which method is the best. Since performance of all MT methods are virtually indistinguishable from each other we also select 5-gram as a representative MT method when comparing with phonetic and similarity methods. }
\label{fig:moses}
\end{figure*}

Figure \ref{fig:moses} shows how we selected the best of MT methods. Even though that for all of our data test folds MT methods produced overlapping confidence bands we can still see that the centroid curve for 5-gram MT methods slightly outperforms other n-gram methods. Therefore, we have selected it to represent MT methods when comparing with phonetic and similarity methods.

\begin{figure*}
  \centering\includegraphics[width=.9\linewidth,height=4in]{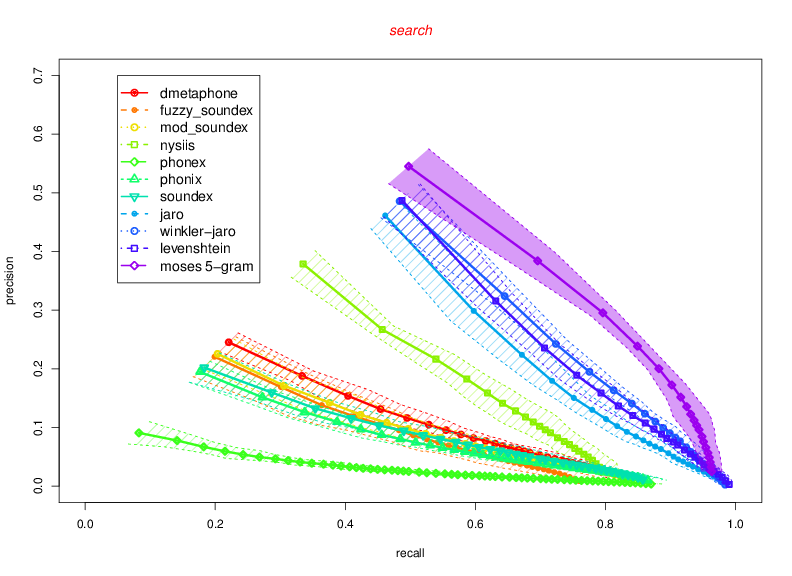}\par
  \caption{``Search'' dataset. MT moses-5-gram method outperforms all other methods and similarity methods generally outperform phonetic methods. The Jaro-Winkler and the Levenshtein confidence intervals overlap slightly with top position of moses-5-gram confidence interval. Similarity measures such as Levenshtein, Jaro and Jaro-Winkler all perform better than phonetic methods but are inferior when comparing them with Machine Translation methods.  NYSIIS appears to significantly outperform other phonetic algorithms.}
  \label{fig:search1140}
\end{figure*}

\begin{figure*}
  \centering\includegraphics[width=.9\linewidth,height=4in]{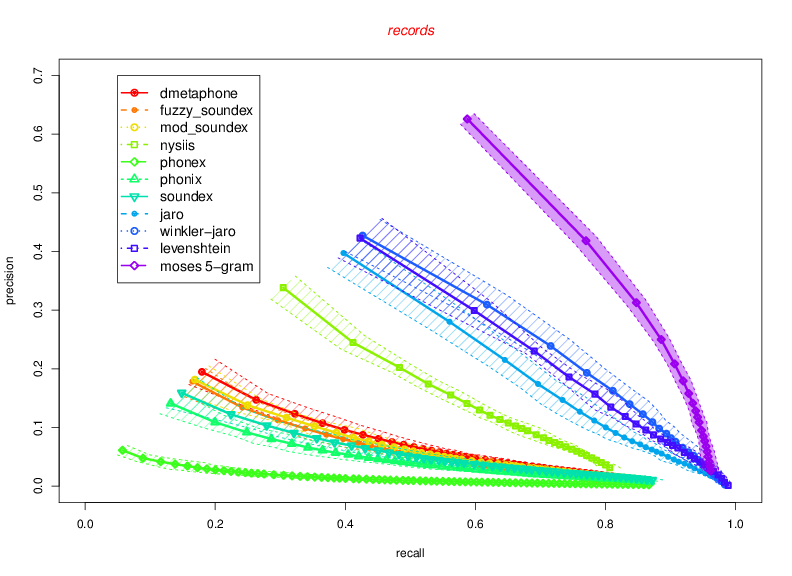}\par
  \caption{``Records'' dataset. MT moses-5-gram method clearly outperforms all other methods. The general order of methods looks similar to the order resulted from running on ``search'' dataset. Similarity measures such as Levenshtein, Jaro and Jaro-Winkler all perform better than phonetic methods but are inferior when comparing them with Machine Translation methods. NYSIIS appears to significantly outperform other phonetic algorithms.}
  \label{fig:records1140}
\end{figure*}

Our main results are shown on Figures \ref{fig:search1140} and \ref{fig:records1140}. Here we present the comparison of all alternative name generating methods on precision-recall plots. It is clear that for both datasets MT methods perform better than all other methods and that similarity methods generally outperform phonetic methods.

\subsection{Implementation details}
We imported our records/tree datasets into CDH4 Cloudera Hadoop and we perform all our filtering using Hive/Python scripts and Java native implementations. The Febrl library \cite {Christen08febrl} implemented by Christen was used for calculating phonetic codes and  string similarity values.


\begin{center}
\begin{table*}[!htb]
    
    \begin{minipage}{.55\linewidth}

        \begin{tabular}{ |l|l|l| }
\hline
Country & number of unique names  \\ [0.4ex] 
\hline
 
England & 9341  \\
Germany & 5679  \\
France & 1233  \\
Ireland & 981  \\
Scotland & 647  \\
Russia & 448  \\
Italy & 426  \\
Switzerland & 377  \\
Norway & 376  \\
Netherlands & 300  \\
Others	& 3779 \\
\hline
\end{tabular}
    \caption{``Records'' dataset
}
    \end{minipage}%
    \begin{minipage}{.55\linewidth}

        \begin{tabular}{ |l|l|l| }
\hline
Country & number of unique names  \\ [0.4ex] 
\hline
 
England & 6690  \\
Germany & 1323  \\
Ireland & 900  \\
France & 631  \\
Scotland & 468  \\
Russia & 241  \\
Italy & 157  \\
Sweden & 109\\
Poland & 90\\
Switzerland & 83  \\
Others & 727 \\
\hline
\end{tabular}
    \caption{``Search'' dataset 
}
    \end{minipage} 
\end{table*}
\end{center}
\section{Discussion and Conclusion}

In this paper we presented a novel way of approaching alternative name spelling generation problem. We utilized a well-known methodology for comparing alternative name spelling methods and presented our results as precision-recall plots which clearly indicate not only that machine translation methods appear to be superior for our datasets to other methods but also show the rankings of other well known methods. We demonstrated our results using a unique dataset from Ancestry.com generated by millions of motivated users who are ``experts'' at labeling the dataset. 

The main conclusion of this work is that machine translation methods that we have employed for finding ranked list of alternative last name spellings far-outperformed all other methods we tried. Our results, also, indicated that the NYSIIS phonetic method significantly outperformed other phonetic algorithms and the Phonex phonetic method did not perform as well on our data. Additionally, Jaro-Winkler similarity method together with the Levenshtein edit distance method performed better than the Jaro method, which was in line with our expectations. On the other hand, we were surprised by how well the NYSIIS method performed compared to other phonetic methods. Our finding regarding phonetic methods performance went against findings reported by Christen in his 2006 paper \cite{christen2006comparison}. However, he was relying on very different dataset and that may explain the differences in our results.

In future work we plan on training our models specifically on training sets composed of name pairs from the same country we plan on testing them against. We also plan on doing more experiments with full names including first names and initials. Additionally, we plan on trying MT methods on geographical locations such as town/village names.

\begin{figure}
\centering
\includegraphics[width=9.2cm]{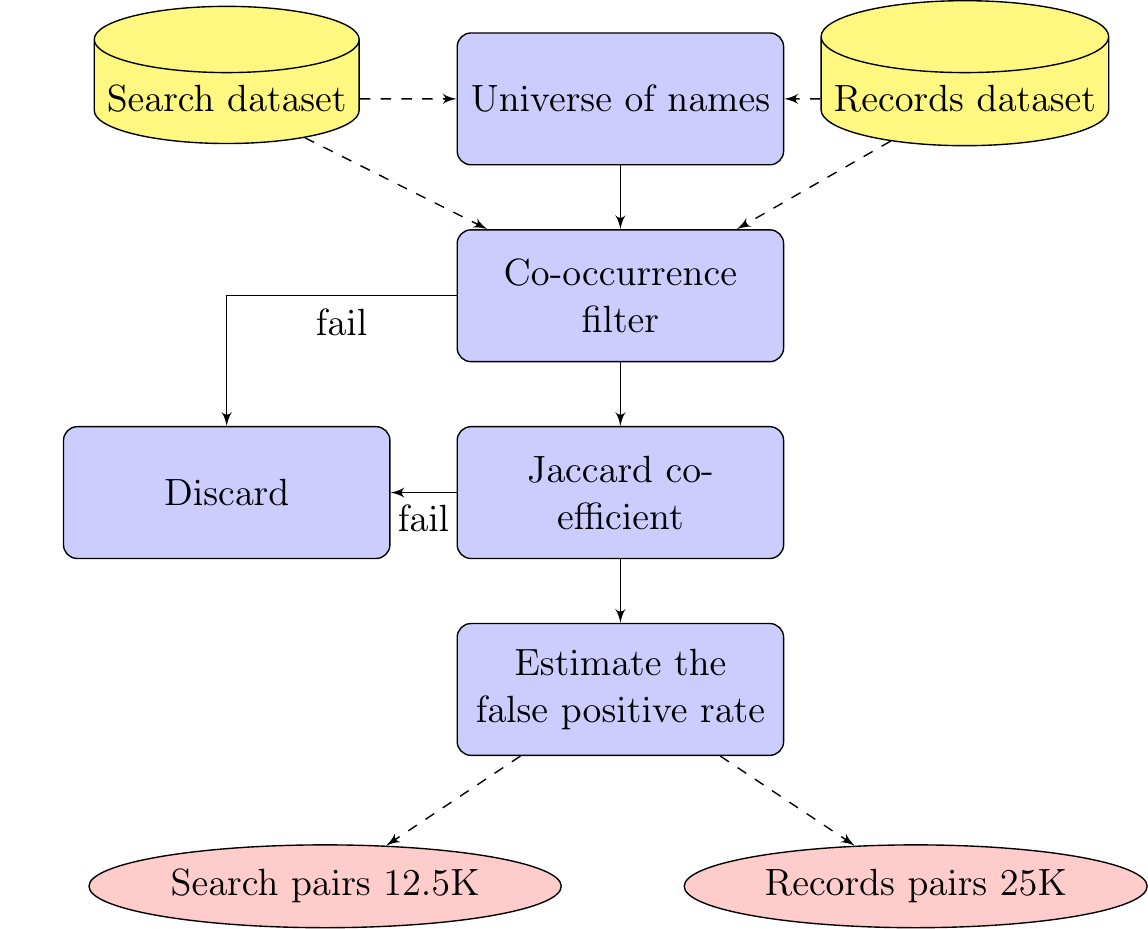}
\caption{Train/Test datasets preparation }
\label{fig:dataprep}
\end{figure}

\clearpage
\bibliography{paper}

\begin{thebibliography}{10}

\bibitem{acom12}
Ancestry.com inc. reports: Q1 2012 financial results.

\bibitem{BhagatH07}
R.~Bhagat and E.~H. Hovy.
\newblock Phonetic models for generating spelling variants.
\newblock In M.~M. Veloso, editor, {\em IJCAI}, pages 1570--1575, 2007.

\bibitem{bilenko03}
M.~Bilenko, R.~Mooney, W.~Cohen, P.~Ravikumar, and S.~Fienberg.
\newblock Adaptive name matching in information integration.
\newblock {\em Intelligent Systems, IEEE}, 18(5):16--23, 2003.

\bibitem{BilenkoKDD03}
M.~Bilenko and R.~J. Mooney.
\newblock Adaptive duplicate detection using learnable string similarity
  measures.
\newblock In {\em In Proceedings of the Ninth ACM SIGKDD International
  Conference on Knowledge Discovery and Data Mining (KDD-2003}, pages 39--48,
  2003.

\bibitem{Bradford13}
R.~B. Bradford.
\newblock Use of latent semantic indexing to identify name variants in large
  data collections.
\newblock In {\em ISI}, pages 27--32, 2013.

\bibitem{Brown1990}
P.~F. Brown, J.~Cocke, S.~A.~D. Pietra, V.~J.~D. Pietra, F.~Jelinek, J.~D.
  Lafferty, R.~L. Mercer, and P.~S. Roossin.
\newblock A statistical approach to machine translation.
\newblock {\em Comput. Linguist.}, 16(2):79--85, June 1990.

\bibitem{Brown93}
P.~F. Brown, V.~J. Pietra, S.~A.~D. Pietra, and R.~L. Mercer.
\newblock The mathematics of statistical machine translation: Parameter
  estimation.
\newblock {\em Computational Linguistics}, 19:263--311, 1993.

\bibitem{christen2006comparison}
P.~Christen.
\newblock A comparison of personal name matching: Techniques and practical
  issues.
\newblock In {\em Data Mining Workshops, 2006. ICDM Workshops 2006. Sixth IEEE
  International Conference on}, pages 290--294. IEEE, 2006.

\bibitem{Christen08febrl}
P.~Christen.
\newblock Febrl -- an open source data cleaning, deduplication and record
  linkage system with a graphical user interface (demonstration session.
\newblock In {\em In ACM International Conference on Knowledge Discovery and
  Data Mining (SIGKDD'08}, pages 1065--1068, 2008.

\bibitem{data:matching}
P.~Christen.
\newblock {\em Data Matching: Concepts and Techniques for Record Linkage,
  Entity Resolution, and Duplicate Detection}.
\newblock Data-centric systems and applications. Springer, 2012.

\bibitem{Cohen03acomparison}
W.~W. Cohen, P.~Ravikumar, and S.~E. Fienberg.
\newblock A comparison of string distance metrics for name-matching tasks.
\newblock pages 73--78, 2003.

\bibitem{dempster1977}
A.~P. Dempster, N.~M. Laird, and D.~B. Rubin.
\newblock Maximum likelihood from incomplete data via the em algorithm.
\newblock {\em Journal of the Royal Statistical Society. Series B
  (Methodological)}, pages 1--38, 1977.

\bibitem{Federico08}
M.~Federico, N.~Bertoldi, and M.~Cettolo.
\newblock {IRSTLM}: an open source toolkit for handling large scale language
  models.
\newblock In {\em INTERSPEECH}, pages 1618--1621. ISCA, 2008.

\bibitem{fellegi1969}
I.~P. Fellegi and A.~B. Sunter.
\newblock A theory for record linkage.
\newblock {\em Journal of the American Statistical Association},
  64(328):1183--1210, 1969.

\bibitem{Hansen05thecma}
N.~Hansen.
\newblock The {CMA} evolution strategy: A tutorial, 2005.

\bibitem{jaro89}
M.~A. Jaro.
\newblock Advances in record-linkage methodology as applied to matching the
  1985 census of {T}ampa, {F}lorida.
\newblock {\em Journal of the American Statistical Association},
  84(406):414--420, 1989.

\bibitem{jurafsky2008speech}
D.~Jurafsky and J.~H. Martin.
\newblock Speech and language processing: An introduction to speech
  recognition.
\newblock {\em Computational Linguistics and Natural Language Processing. 2nd
  Edn., Prentice Hall, ISBN}, 10(0131873210):794--800, 2008.

\bibitem{kernighan1990spelling}
M.~D. Kernighan, K.~W. Church, and W.~A. Gale.
\newblock A spelling correction program based on a noisy channel model.
\newblock In {\em Proceedings of the 13th conference on Computational
  linguistics-Volume 2}, pages 205--210. Association for Computational
  Linguistics, 1990.

\bibitem{Koehn2007}
P.~Koehn, H.~Hoang, A.~Birch, C.~Callison-Burch, M.~Federico, N.~Bertoldi,
  B.~Cowan, W.~Shen, C.~Moran, R.~Zens, C.~Dyer, O.~Bojar, A.~Constantin, and
  E.~Herbst.
\newblock Moses: Open source toolkit for statistical machine translation.
\newblock In {\em Proceedings of the 45th Annual Meeting of the ACL on
  Interactive Poster and Demonstration Sessions}, ACL '07, pages 177--180,
  Stroudsburg, PA, USA, 2007. Association for Computational Linguistics.

\bibitem{Levenshtein66}
V.~I. Levenshtein.
\newblock Binary codes capable of correcting deletions, insertions and
  reversals.
\newblock {\em Soviet Physics Doklady}, 10(8):707--710, February 1966.

\bibitem{Macskassy04}
S.~A. Macskassy and F.~J. Provost.
\newblock {Confidence Bands for ROC Curves: Methods and an Empirical Study}.
\newblock In {\em ROC Analysis in Artificial Intelligence}, pages 61--70, 2004.

\bibitem{McCallumBP05}
A.~McCallum, K.~Bellare, and F.~C.~N. Pereira.
\newblock A conditional random field for discriminatively-trained finite-state
  string edit distance.
\newblock In {\em UAI}, pages 388--395. AUAI Press, 2005.

\bibitem{MoreauYC08}
E.~Moreau, F.~Yvon, and O.~Cappe.
\newblock Robust similarity measures for named entities matching.
\newblock In D.~Scott and H.~Uszkoreit, editors, {\em COLING}, pages 593--600,
  2008.

\bibitem{och03:asc}
F.~J. Och and H.~Ney.
\newblock A systematic comparison of various statistical alignment models.
\newblock {\em Computational Linguistics}, 29(1):19--51, 2003.

\bibitem{Pfeifer96}
U.~Pfeifer, T.~Poersch, and N.~Fuhr.
\newblock Retrieval effectiveness of proper name search methods.
\newblock In {\em Information Processing and Management}, pages 667--679, 1996.

\bibitem{philips90}
L.~Philips.
\newblock Hanging on the metaphone.
\newblock {\em Computer Language Magazine}, 7(12):39--44, December 1990.

\bibitem{Ristad96}
E.~S. Ristad and P.~N. Yianilos.
\newblock Learning string edit distance.
\newblock {\em CoRR}, cmp-lg/9610005, 1996.

\bibitem{Russell18}
R.~Russell.
\newblock Soundex.
\newblock {\em U.S. Patent 1,261,167}, 04 1918.

\bibitem{shannon48}
C.~Shannon.
\newblock A mathematical theory of communication.
\newblock {\em Bell System Technical Journal}, 27:379--423, 623--656, July,
  October 1948.

\bibitem{Taft70}
R.~L. Taft.
\newblock Name search techniques.
\newblock Technical Report Special Report No. 1, New York State Identification
  and Intelligence System, Albany, NY, February 1970.

\bibitem{winkler90}
W.~E. Winkler.
\newblock String comparator metrics and enhanced decision rules in the
  fellegi-sunter model of record linkage.
\newblock In {\em Proceedings of the Section on Survey Research}, pages
  354--359, 1990.

\bibitem{Winkler06}
W.~E. Winkler.
\newblock Overview of record linkage and current research directions.
\newblock Technical report, Bureau of the Census, 2006.

\end{thebibliography}
\end{document}